\documentclass[aps,twocolumn,floats,pre]{revtex4}
\usepackage{graphics,graphicx,epsfig}
\usepackage{amssymb,color}
\usepackage{epsf,epstopdf,wrapfig}

\begin{document}

\title{From modes to movement  in the  behavior of {\em C. elegans}}

\author{Greg J Stephens, Bethany Johnson-Kerner, William Bialek and William S Ryu\footnote{Present address:  Department of Physics, Banting and Best Department of Medical Research,  University of Toronto, Toronto M5S 1A7, Ontario, Canada} }

\affiliation{Joseph Henry Laboratories of Physics and
Lewis--Sigler Institute for Integrative Genomics,
Princeton University, Princeton, New Jersey 08544 USA}

\begin{abstract}
Organisms move through the world by changing their shape, and here we explore the mapping from shape space to movements in the nematode {\em C. elegans} as it crawls on a planar agar surface.  
We characterize the statistics of the trajectories through the correlation functions of the orientation angular velocity, orientation angle and the mean-squared displacement, and we find that the loss of
orientational memory has significant contributions from both abrupt, large amplitude turning events and the continuous dynamics between these events.  Further, we demonstrate long-time 
persistence of orientational memory in the intervals between abrupt turns.  Building on recent work demonstrating that {\em C. elegans} movements are restricted to a low-dimensional shape space, we construct a map from the dynamics in this shape space to the trajectory of the worm along the agar.  We use this connection to illustrate that changes in the continuous dynamics reveal subtle differences in movement strategy that occur among mutants defective in two classes of dopamine receptors.  
\end{abstract}

\date{\today}

\maketitle

\section{Introduction}

From the swimming motions of {\em E. coli}  \cite{berg_book} to the mobility of human populations \cite{brockmann_et_al_06}, the way in which organisms move through the world profoundly influences their  experience. Ultimately, these strategies of movement change the chances for survival and reproduction, and thus are subject to natural selection.  Crucially, whether a movement is adaptive depends on its relation to the outside world, but the organism has control only over its internal states.  From this internal point of view, movement is not translation or rotation of the body relative to a fixed external coordinate system, but rather transformations of the organism's shape as measured in its own intrinsic coordinates.   

In general, the connection between transformations in shape space and movement through the world is complicated.  There is a long tradition of work which tries to make this connection through analytic approximations of the equations describing the mechanics of the organism's interaction with the outside world.  This approach is perhaps best developed for swimming and flying organisms \cite{childress_81,lighthill_87}, and there are particularly elegant results in the limit of swimming at low Reynolds' number \cite{shapere+wilczek_87,shapere+wilczek_89a,shapere+wilczek_89b}.  All of these methods depend on some small parameter in the physical interaction between the organism and its environment.  A very different possibility for simplifying the relation between shapes and movement arises if the space of shapes itself is limited.    In several systems, the potentially high dimensional space of shapes or movements is not sampled uniformly under natural conditions, so that one can recognize a lower dimensional manifold that fully describes the system \cite{avella+bizzi_98,santello+al_98,sanger_00,osborne+al_05}.  In these cases it is possible to ask empirically how motions on this low dimensional manifold map into movements relative to the outside world.

The motion of the nematode {\em C. elegans} provides an example of this dimensionality reduction.  In previous work, we found that the shapes taken on by the worm's body are well-approximated by a four dimensional space spanned by elementary shapes or `eigenworms'  \cite{stephens+al_08}.  Here we connect the dynamics in this low dimensional space of shapes to the trajectories of worms as they crawl on an agar plate, the conventional experimental setup for studying worm behavior \cite{gray+al_05, ramot+al_08, biron+al+08,faumont+lockery_05}.  In the process, we offer a new analysis of the trajectories themselves, and show how the intrinsic shape dynamics gives us a more refined tool for the analysis of mutant locomotory behavior.

\section{Center of mass trajectories}

\begin{figure}[htb]
\begin{center}
\includegraphics[width=0.9\columnwidth]{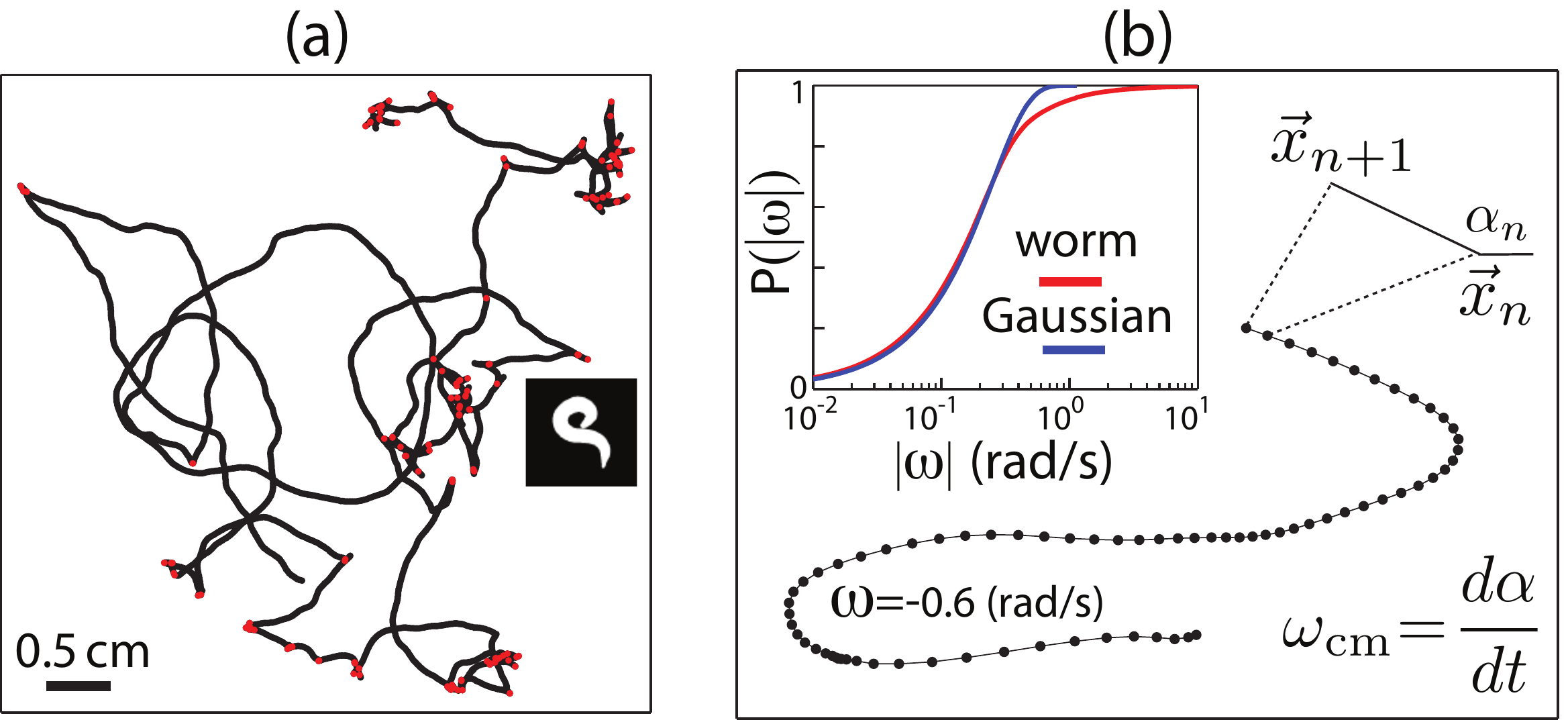}
\end{center}
\caption{Centroid trajectories.  (a) A typical center of mass trajectory on the agar plate.  The path includes both abrupt orientation changes (red) in which the worm shape is deeply bent (inset), and continuously curving segments. (b) Definition of the local orientation angle $\alpha(t)$ and the local orientation velocity $\omega(t)={d\alpha(t)}/{dt}$.  The inset shows  the cumulative distribution of orientation velocities (red) contrasted with a Gaussian distribution that has the same variance (blue).  The centroid time series $\{x_{{\rm cm}}(t),y_{{\rm cm}}(t)\}$ was filtered with a third-order polynomial in a running window 11 frames in length and  $\alpha_{{\rm cm}}$ and  $\omega_{{\rm cm}}$ were built using this filter. }
\label{cm-traj}
\end{figure}

To understand how the motion of the worm is determined by its internal dynamics in shape space, we first characterize the motion itself.  In Fig \ref{cm-traj}a, we show an example of the worm's trajectory, defined as the centroid of the black and white body image.  These data are collected as in Ref \cite{stephens+al_08}, using tracking microscopy, and what is shown is one example from a data set that includes  $N=33$ wild-type worms, each tracked for 35 minutes.  The worms were transfered to the agar plate using a platinum worm pick and we
excluded the first 400 seconds of each tracking run to avoid any influences from this mechanical stimulation.

The trajectory consists of gently curving segments, interrupted by sharp turns, often associated with characteristically curved body shapes; these discrete events are known as $\Omega$--turns.  It is therefore tempting to think of these trajectories as being approximately like those of {\em E. coli}, consisting of long, relatively straight runs punctuated by tumbles, which randomly reorient the cell \cite{berg_93}.  Indeed, variations of this model in which trajectories are segmented into discrete `runs'  have long been used in study of {\em C. elegans} \cite {lockery_99, gray+al_05} and in the analysis of movements for a wide variety of organisms  \cite{edwards+al_07,bartumeus+al_03}.

Trajectories are characterized by the speed  of motion, $v_{\rm cm}=|\dot{\vec{x}}|$,   the local tangent angle, $\alpha(t)$, and the rate of local curvature  $\omega = d\alpha/dt$ at each moment in time (Fig \ref{cm-traj}b).  The standard deviation of the curvature is  $\delta\omega_{\rm rms} =0.223 \pm 0.001\,{\rm rad/sec}$, but $\rho(\omega)$ has long tails, as shown by comparing the cumulative distribution $P(|\omega|)$ to a Gaussian distribution with the same variance (inset to Fig \ref{cm-traj}b).  Excursions to the large amplitude tails correspond to abrupt reorientation events and are colored red in Fig 1a.

\begin{figure}[b]
\begin{center}
\includegraphics[width=\columnwidth]{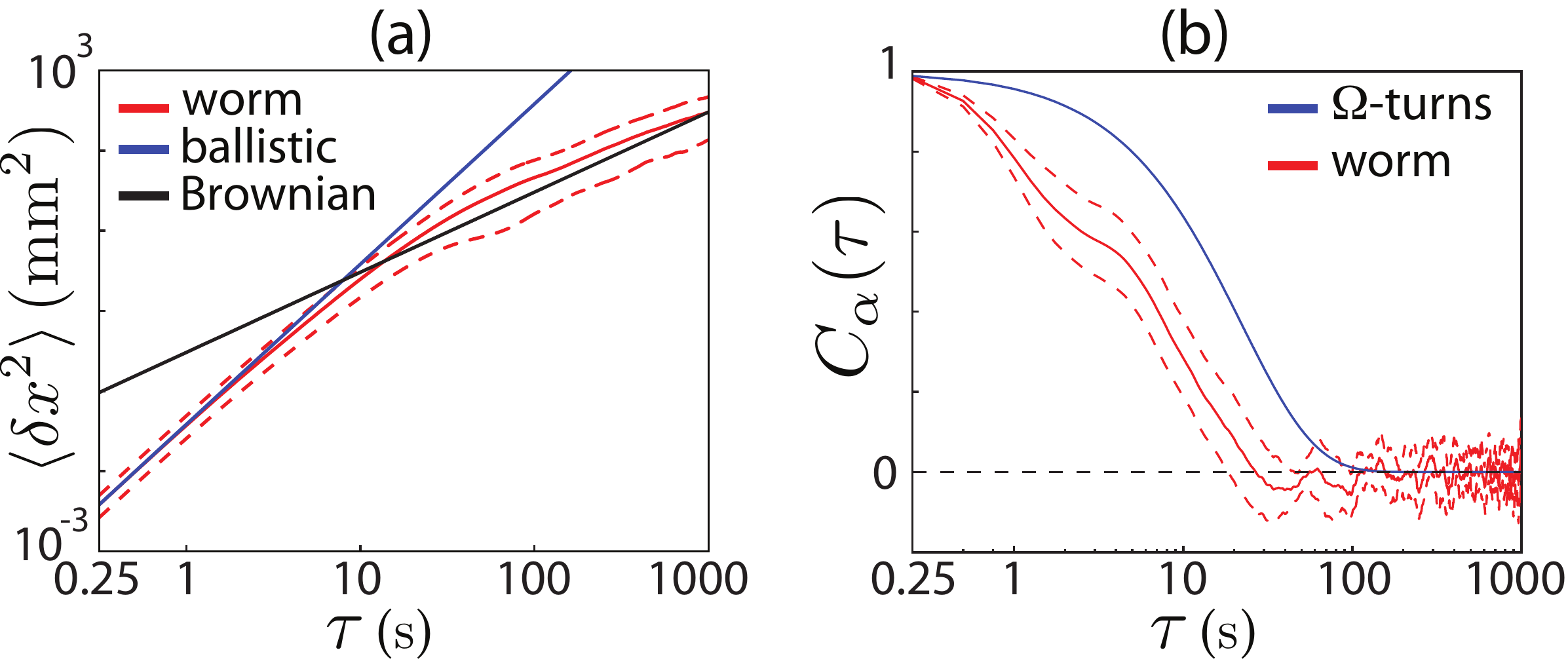}
\end{center}
\caption{Trajectory correlations. (a) Mean--square displacement as a function of time along the  trajectory [Eq (\ref{Deltax})].  At short times we see
ballistic ($\delta x \propto \tau$) behavior, crossing over to diffusion 
[$(\delta x)^2 \propto \tau$] at long times.
(b)  The orientation angle correlation function [Eq (\ref{Calpha})] for worm motion (red), contrasted  with a simple ``run and tumble'' model (blue) in which reorientation events occur randomly with rate $r_{\rm \Omega}=0.045\, {\rm Hz}$, corresponding to the rate of $\Omega$--turns   \cite{lockery_99, gray+al_05}.}
\label{cm-cor}
\end{figure}

If worms lose orientational memory  then their movements will look  diffusive at long times.  One signature of such behavior is contained in the mean--square distance between two points on the trajectory, 
\begin{equation}
\langle (\delta x)^2 \rangle_\tau \equiv \langle | \vec x (t+\tau ) - \vec x(t) |^2 \rangle,
\label{Deltax}
\end{equation}
which for diffusion should grow linearly with the time $\tau$.  In Fig \ref{cm-cor}a, we see that this is what happens for times longer than $\tau \sim 10\,{\rm s}$. At shorter times, the mean--square displacement grows as the square of the time difference, which corresponds to ballistic motion at a fixed velocity. This suggests that the time scale for worms to lose directional memory is $\sim 10\,{\rm s}$, and this can be seen directly from the correlation function for the local tangent angle,
\begin{equation}
C_\alpha(\tau ) \equiv \langle\cos[ \alpha(t+\tau ) -\alpha(t)]\rangle ,
\label{Calpha}
\end{equation}
as shown in Fig \ref{cm-cor}b. 

A sequence of uncorrelated runs and tumbles translates into a mathematical model in which the orientation angle $\alpha$ is, from time to time, completely randomized by discrete events.  In {\em E. coli} these events are the tumbles, and it is natural to think that in {\em C. elegans} these events are the $\Omega$--turns.   If turns occur randomly at rate $r_{{\rm turn}}$, and generate completely new, uncorrelated directions of movement, then the correlation function for direction will decay  as 
\begin{equation}
C_\alpha(\tau )  = \exp\left(  - r_{{\rm turn}} |\tau |\right) .
\end{equation}
But since $r_{{\rm turn}}$  typically is less than two per minute \cite{lockery_99,gray+al_05}, $\Omega$--turns alone can't explain the shorter time of decay of the directional correlations, as seen quantitatively in Fig \ref{cm-traj}d.  Thus,  changes in the orientation angle $\alpha$ occurring in between $\Omega$--turns must play a key role.

\begin{figure}[htb]
\begin{center}
\includegraphics[width=0.85\columnwidth]{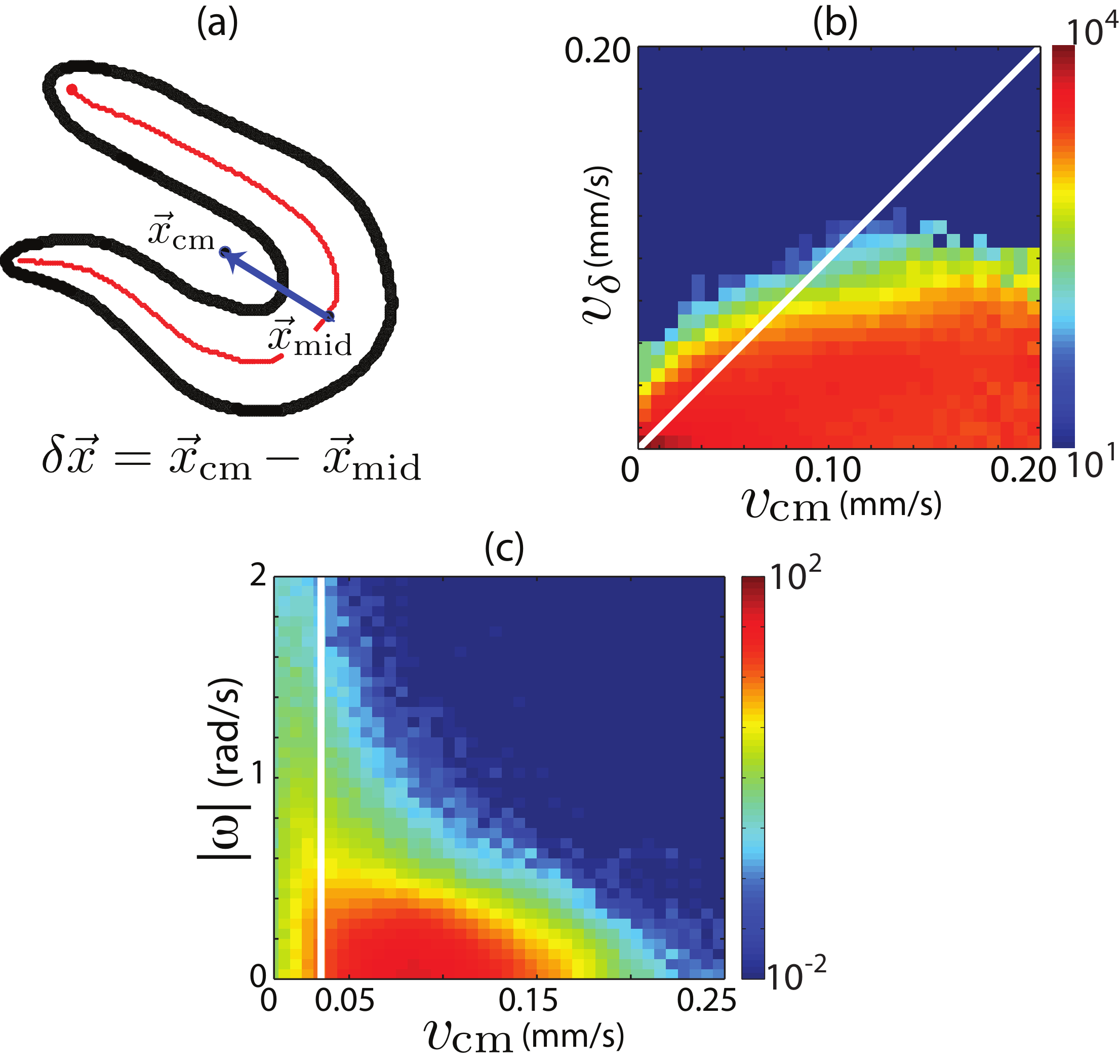}
\end{center}
\caption{The reorientation of directional heading.  (a)  The displacement vector $\delta \vec{x}=\vec{x}_{{\rm cm}}-\vec{x}_{{\rm mid}}$. 
(b) The conditional probability density  $\rho(v_{\delta} | v_{\rm cm})$.  At low speeds, the centroid velocity mixes changes of the body shape with changes in the directional heading.  (c) Low centroid speeds are also associated with large centroid orientation changes as shown
by the conditional distribution $\rho(\omega | v_{\rm cm})$.  The solid white line denotes the speed threshold at which we separate continuous intervals from discrete events. }
\label{cm-heading}
\end{figure}

To find an objective but tractable definition of the discrete events that interrupt the trajectory, we examine
the approximations made in drawing the  trajectory itself, as in Fig \ref{cm-traj}.  Following general practice, we generated trajectories by tracking the center of mass of the worm's silhouette, ${\vec x}_{\rm cm}(t)$.   An alternative would be to follow the midpoint of the worm as indicated in Fig \ref{cm-heading}a, generating a trajectory ${\vec x}_{\rm mid}(t)$.  When the worm moves smoothly and doesn't change shape dramatically, these different definitions of trajectory produce nearly identical results.  But, during those moments when the worm pauses and reorients, these trajectories diverge.  This gives us a natural way to identify abrupt reorientation events. 

We look at the difference between the center of mass and the midpoint trajectories, $\delta \vec{x}=\vec{x}_{{\rm cm}}-\vec{x}_{{\rm mid}}$ and define the difference speed $v_{\delta}=\delta \dot x$.  From the conditional density $\rho(v_{\delta} | v _{\rm cm})$, shown in Fig \ref{cm-heading}b, we can see that  this difference speed $v_\delta$ is typically much smaller than the center of mass speed $v_{\rm cm}$.  But at small $v_{\rm cm}$ there is substantial probability that $v_\delta>v_{\rm cm}$ and when this occurs the centroid orientation is no longer an accurate measure of the worm's heading.  At low speeds there is also a large variance in local curvature, which can be seen in the conditional density $\rho(\omega_{\rm cm} | v_{\rm cm})$, Fig \ref{cm-heading}c. 
Thus our segmentation based on $v_{\rm cm}$ is similar to those based on large values of $\omega_{\rm cm}$.
To simultaneously isolate the largest amplitude turns and maintain the validity of the centroid orientation within intervals we choose $v_{\rm thresh}=0.035\,{\rm mm/s}$.  Since lower centroid speeds are associated with curvier shapes, the choice
of a threshold in $v_{\rm cm}$ amounts to tolerating a fraction of  frames  with $v_\delta>v_{\rm thresh}$ and this fraction is 25\% at our threshold.    Most of the time $v_{\rm cm}$ is larger and the fraction substantially smaller.

\begin{figure}[htb]
\begin{center}
\includegraphics[width=0.8\columnwidth]{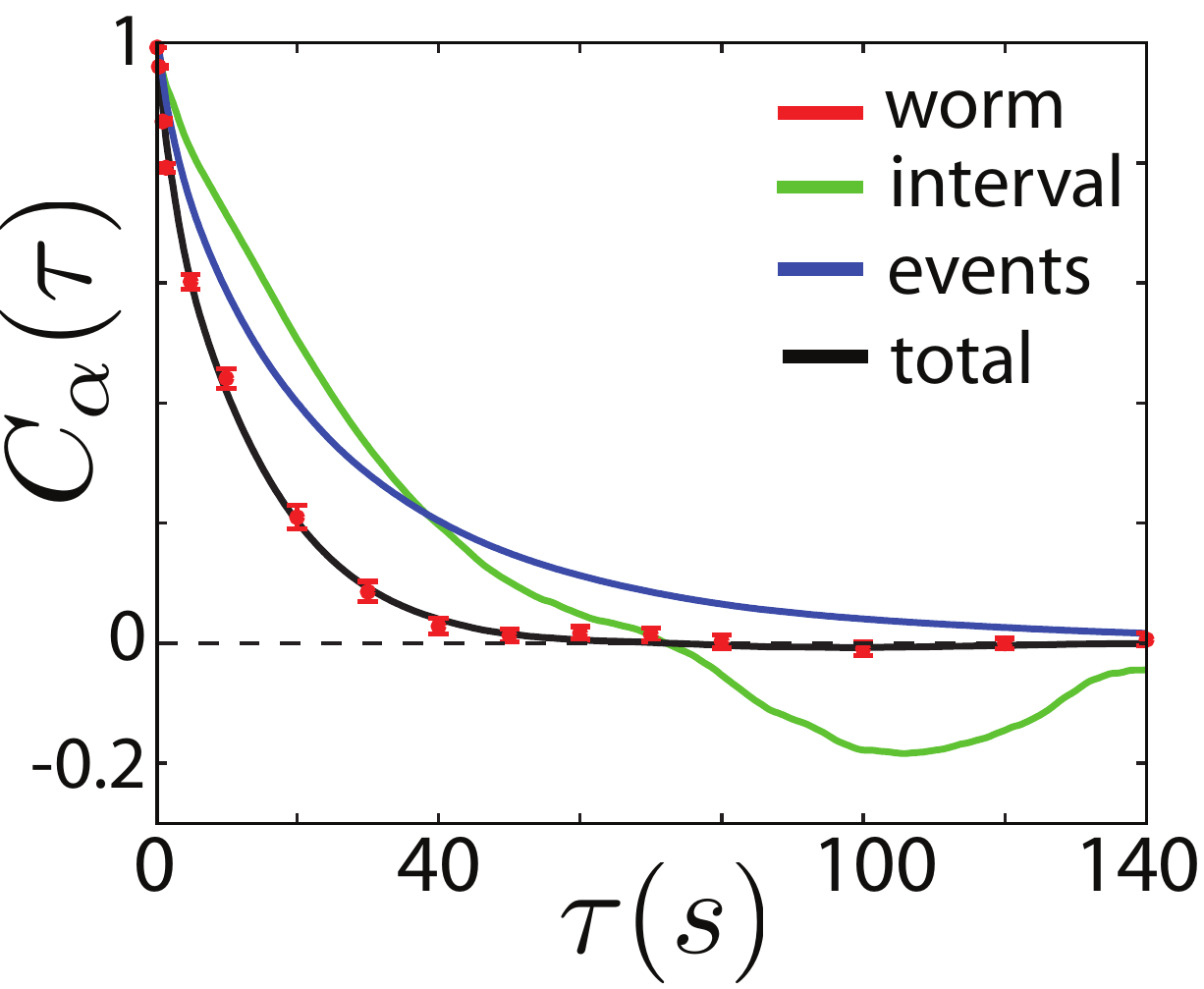}
\end{center}
\caption{The loss of orientation memory is  captured by independent contributions from abrupt reorientation events and continuous dynamics between these events. 
The orientation correlation function $C_{\alpha}^{\rm interval}(\tau)$ 
computed by averaging over all delays within the same interval is shown in green. The probability $P(\tau )$ that two times separated by $\tau$ are within the same interval is shown in blue.
The predicted orientation correlation function from Eq (\ref{eq-alphaCorTotal}) is shown in black,  compared with the data, shown as red points with standard errors of the mean.}
\label{alphaCor_total}
\end{figure}

Using the intervals between reorientation events as defined above, we can disentangle the contributions of continuous and discrete reorientation.   We compute the orientation correlation function $C_{\alpha}(\tau)$  within the intervals, and the result, $C_{\alpha}^{\rm interval} (\tau)$, is shown as the green curve of Fig.~{\ref{alphaCor_total}. Removing the intervening abrupt reorientation events reveals angular correlations which decay on a much longer time scale.  Furthermore we observe {\it anti}--correlations which indicate a marked departure from ordinary orientational diffusion.  

As an approximation, we assume that the continuous dynamics of orientation within arcing segments of the trajectory are independent of the discrete reorientation events, and that the reorientation events are instataneous.  Then the full orientation correlation function is the product of two terms,
\begin{equation}
C_{\alpha}^{\rm total}(\tau)=C_{\alpha}^{\rm interval} (\tau)\times P(\tau),
\label{eq-alphaCorTotal}
\end{equation}
where $P(\tau)$ is the measured probability that two points separated by time $\tau$ come from the same interval.   Figure \ref{alphaCor_total} shows that this is a excellent approximation, with no adjustable parameters.

\section{Modes and motions}

Following the center of mass of an organism is a convenient but approximate measure of its movement strategy.  A fuller, and organism centered, representation of locomotion is contained in the shape of the worm itself.  In previous work \cite{stephens+al_08} we found that the space of shapes of {\em C. elegans} during free locomotion is low-dimensional with four
principal dimensions (eigenworms) capturing approximately $95\%$ of the variance of the space of shapes.  
A summary of these results is shown in Fig \ref{eigenworms}.

\begin{figure}[htb]
\includegraphics[width=\columnwidth]{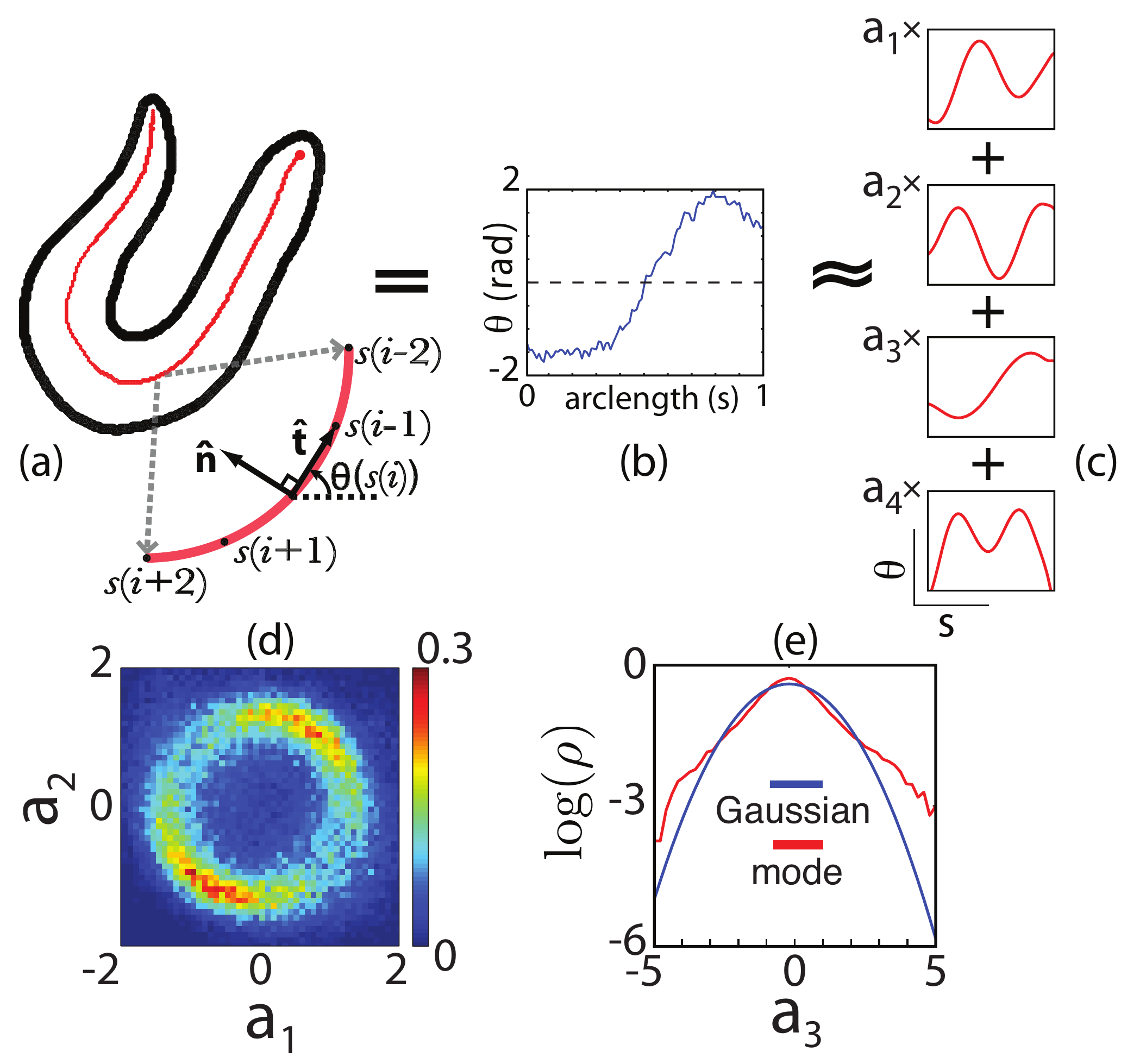}
\caption{Low dimensional description of worm shape, following Ref \cite{stephens+al_08}. (a) Each raw image of the worm   is processed
by passing a curve through the center of the body; the red circle marks the head.  Arc length $s$ along this curve is normalized, $0 \leq s < 1$, and
we define the tangent ${\bf\hat{t}}(s)=\frac{d\theta}{ds}$ to the centerline curve. (b) We rotate all images so that $\langle \theta \rangle $ is zero and thus $\theta(s)$ provides a description of the wormÕs shape that is intrinsic to the worm itself.  (c) We decompose each shape $\theta (s)$ into contributions from the leading four eigenmodes, which capture 95\% of the variances in shape space, and the amplitudes of fluctuation along each mode are normalized to unit variance, $\langle a_i^2\rangle=1$.
(d) Fluctuations along the first two modes correspond to an oscillation, or nearly circular orbit in this projection of the shape space. 
(e)   Fluctuations along the third mode show strongly non--Gaussian statistics.
 The first two modes drive the propulsive wave along the body, the third mode makes the dominant  contribution to turning, and the fourth mode (not shown)  describes localized fluctuations of the head and tail.}
\label{eigenworms}
\end{figure}

\begin{figure}[htb]
\begin{center}
\includegraphics[width=\columnwidth]{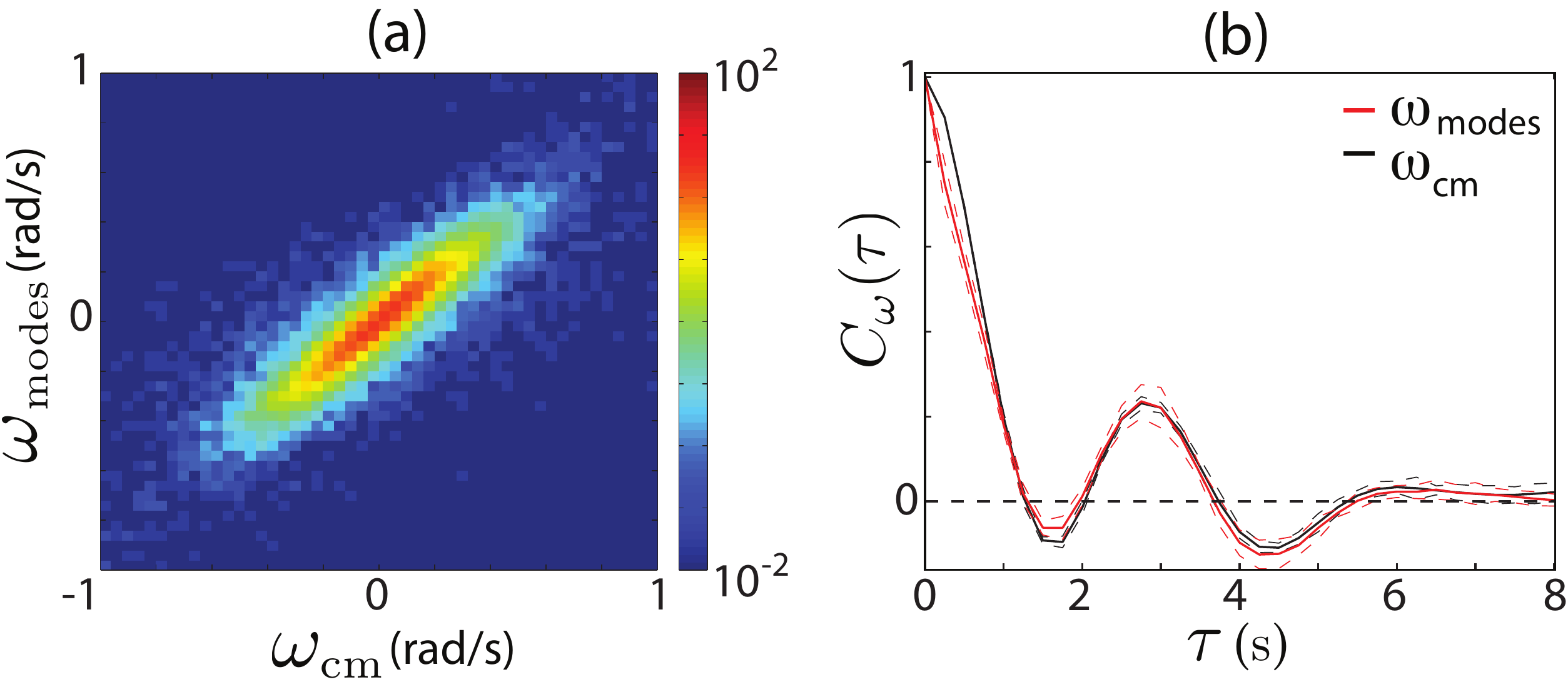}
\end{center}
\caption{The intrinsic dynamics of worm shape capture the extrinsic motions of foraging trajectories.   (a)  We construct an eigenworm model of the orientation velocity (Eq.~\ref{eq:omegaModes}) and show the joint distribution $\rho(\omega_{\rm cm},\omega_{\rm modes})$.  (b) The orientation velocity two-point correlation function for the eigenworm model (blue) and the centroid motion (red).}  
\label{fig:omega_modes}
\end{figure}

To construct the connection between modes and movement we confine our analysis to continuous 
intervals, within which  $\omega_{\rm cm}$ is a faithful measure of the worm's orientational velocity.  We then try to construct a map from the amplitudes of motion along the four intrinsic shape modes into $\omega (t)$.  The simplest model is a linear one,
\begin{equation}
\omega_{\rm modes}(t)=\sum_{{\rm i}=1}^4\beta_{\rm i}a_{\rm i}(t) ,
\end{equation}
and by adjusting the coefficients $\beta_{\rm i}$ we can capture $55\%$ of the variance in $\omega$.   More generally we can try a nonlinear relationship, and allow that the current curvature of the trajectory also depends on the time derivatives of the shape.  To do this, we introduce twelve variables,
\begin{eqnarray}
x_{{\rm i} = 1:4} &=& a_{1:4}(t), \\
x_{{\rm i} = 5:8} &=& {{da_{1:4}(t)}\over{dt}} ,\\
x_{{\rm i} = 9:12} &=& {{d^2a_{1:4}(t)}\over{dt^2}},
\end{eqnarray}
and construct an estimate of the curvature as
\begin{equation}
\omega_{modes}=\sum_{\rm i} \gamma_{\rm i} x_{\rm i} +
\sum_{\rm ij} \gamma_{\rm ij} x_{\rm i}x_{\rm j} +
\sum_{\rm ijk} \gamma_{\rm ijk} x_{\rm i}x_{\rm j}x_{\rm k}.
\label{eq:omegaModes}
\end{equation}
As shown in Fig \ref{fig:omega_modes}a, this approach captures more than $80\%$ of the variance in the curvature.  More importantly, we also capture the dynamics of the curvature, as shown through the correlation function
\begin{equation}
C_\omega (\tau ) \equiv {1\over{\langle \omega^2\rangle}} \langle \omega(t+\tau ) \omega (t)\rangle 
\end{equation}
in Fig \ref{fig:omega_modes}b.

The curvature is dominated by contributions from the amplitude of the third mode, which plausibly is connected to bending of the worm's body \cite{stephens+al_08}. We have shown previously that the instantaneous speed of the worm $v_{\rm cm}(t) \propto \dot\phi (t)$, where the phase $\phi$ as the angle in the plane $\{a_1, a_2\}$ (cf Fig \ref{eigenworms}d). Taken together, then, we can map from the modes to the speed and local curvature of the center of mass trajectory, so we can completely reconstruct the worm's movements from its shape as a function of time.

\section{Mutants and adaptation}

As with all organisms, {\em C. elegans} behavior is modulated by the organism's experience.  As an example, the rate of $\Omega$--turns decreases systematically with time away from food \cite{gray+al_05}, as well as changing in response to thermal \cite{ryu+samuel_02} and chemical \cite{lockery_99} stimuli.  It is also known that dopamine  plays a significant role in these behaviors \cite{chase+al_04, hills+al_04}.  Here we use the analytic tools developed above to characterize both adaptation and the dopamine mutants {\em dop-2 (vs105)} and {\em dop-3 (vs106)}.

To make connections with earlier work, we follow Ref \cite{lockery_99} and identify $\Omega$--turns as moments when $\omega_{\rm cm} > 0.87 \,{\rm rad/s}$.  With this threshold, we can measure the cumulative distribution of inter--turn intervals, or equivalently the probability that a turn has not yet occurred after a time $\tau$, and this is shown in Fig \ref{fig:Omega}.  During the course of our experiments the worms spend 40 min away from food; allowing for an initial adjustment to being placed on the the plate we divide the last
28.3 min into three equal epochs to search for  adaptation to the environment.   We see that the interval distributions vary systematically with time.  As reported previously, the mean time between turns gets longer as the worms spend more time away from food.  In more detail, we see that the distribution has two components, 
\begin{equation}
P(\tau)=a e^{-r_{\rm short} \tau} +(1-a) e^{-r_{\rm long} \tau},
\label{Pfit}
\end{equation}
and only the slower component $r_{\rm long}$ contributes to the lengthening of the times between turns.  Repeating the analysis for the dopamine mutants, we again find that the short time behavior of the distribution, summarized by $r_{\rm short}$ is unchanging, while the long time behavior varies  across time (Table 1).  We also note that  differences in $\Omega$--turns between the dopamine mutants are statistically insignificant though both mutants turn more frequently than wild-type worms.

 \begin{figure}[htb]
\begin{center}
\includegraphics[width=0.7\columnwidth,keepaspectratio=true]{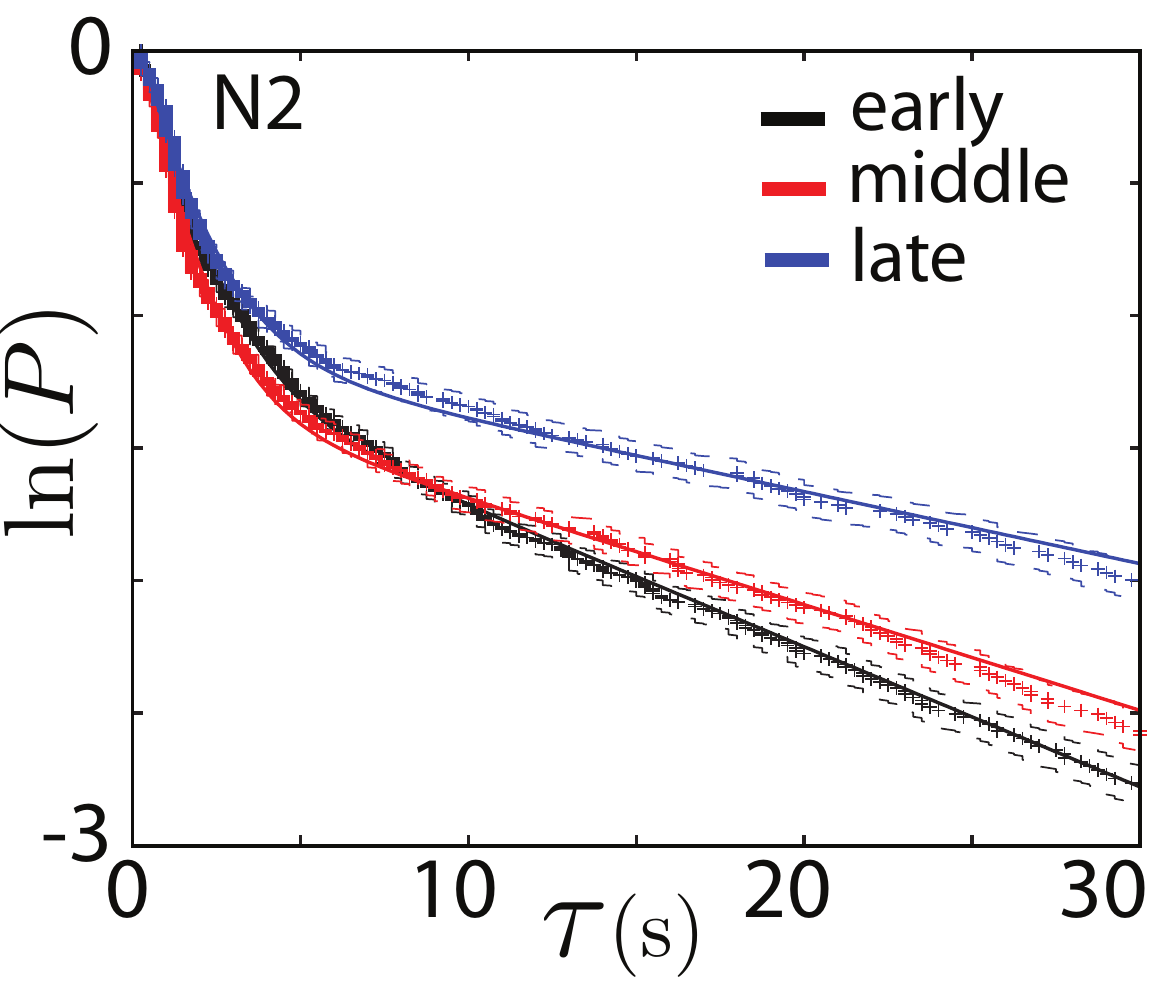}
\end{center}
\caption{The cumulative distribution of times between $\Omega$--turns in wild--type worms.  $\Omega$-events are defined as described in the text, and dashed lines denote bootstrap standard deviations.} 
\label{fig:Omega}
\end{figure}

\begin{table}[htb]
\begin{center}
\begin{tabular}{|c |c|c|c|c|c|c|} \hline
& \multicolumn{3} {|c|} {$r_{\rm short}$ (Hz,\, $\pm 0.02$)} & \multicolumn{3} {|c|} {$r_{\rm long}$ (Hz,\, $\pm 0.002$)} \\ \hline
genotype & early & middle & late  &  early & middle & late \\ \hline
N2 & 0.63 & 0.76 & 0.63  &0.053 & 0.040 & 0.027 \\ \hline
{\em dop-2} & 0.68 & 0.67 & 0.68  & 0.072 &0.048 & 0.044 \\ \hline
{\em dop-3} & 0.65 &0.66 &0.60 & 0.070 & 0.046 & 0.042 \\ \hline
\end{tabular}
\caption{Properties of the inter--turn interval distribution.  Data as in Fig \ref{fig:Omega} were collected also for $N=50$ dopamine mutants, each also observed for 35 minutes, and all the data were fit to Eq (\ref{Pfit}).  Results are shown for the two rates, $r_{\rm short}$ and $r_{\rm long}$, that define the time scales for turning.  We note that   $r_{\rm short}=0.66 \pm 0.04$\,Hz across all epochs and genetic variants, and that differences in $\Omega$--turns between the dopamine mutants are statistically insignificant.}

\end{center}
\label{default}
\end{table}

Our analysis above shows that continuous re--orientation in between turns is a significant component of the worm's motion, and that this behavior is driven by the dynamics in shape space.  Indeed, although the statistics of turning are the same for the two different mutants, Fig \ref{fig:mut} shows that the dynamics along the different modes are in fact quite different.  Along modes 1 and 2 (which form a quadrature pair), {\em dop-2} is similar to the wild type, but {\em dop-3} exhibits a faster oscillation.  Recalling that this corresponds to  the undulatory wave along the body  \cite{stephens+al_08}, we predict that the {\em dop-3} mutant should move more quickly, and this is observed when we track the center of mass motions: mean speeds of the three variants are $v_{\rm cm}=[0.076, 0.082 ,0.121] \pm 0.001$\,mm/s for N2, {\em dop-2} and {\em dop-3} respectively.  The mode dynamics also combine though Eq (\ref{eq:omegaModes}) to produce different turning dynamics  and, in particular,  {\em dop-3} animals make longer-lasting turns which result in curvier trajectories.  While these pheonotypic differences are subtle and have not been reported before, they are immediately apparent in the dynamics in shape space.   Taken together we find continuous motion, including gradual re--orientation, and discrete turning behaviors are under independent genetic control.

\begin{figure}[htb]
\begin{center}
\includegraphics[width=\columnwidth,keepaspectratio=true]{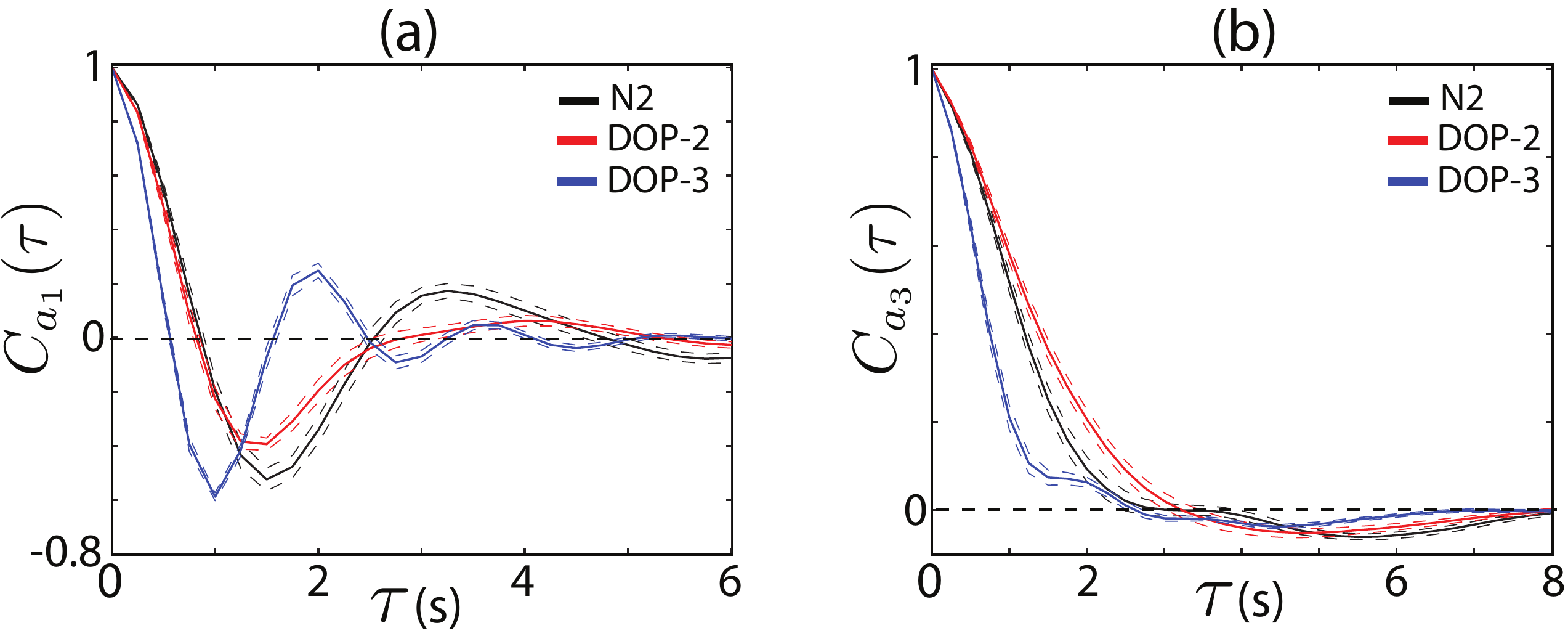}
\end{center}
\caption{Correlation functions for the shape modes $a_1(t)$ and $a_3(t)$, comparing the wild type $N2$ with the two mutant strains {\em dop-2} and {\em dop-3}. Shorter timescales for the {\em dop-3} mutants results in faster centroid speed and curvier trajectories.
Dashed lines denote bootstrap standard deviations.}
\label{fig:mut}
\end{figure}

\section{Discussion}

It is tempting to think of behavior as a sequence of discrete actions, each taken in consequence of a specific decision.  In the simplest cases, such as the running and tumbling of {\em E. coli}, we can see these discrete events reflected directly in the trajectory of the organism's center of mass motion \cite{berg_book}.  In contrast, we find that, for {\em C. elegans}, discrete behaviors are roughly only half of the story: the exploratory trajectories of the worm get approximately equal contributions from discrete turning events and from continuous re--orientational motions in between the turns.   Further, we can trace these continuous motions back to the underlying dynamics in the space of body shapes.  Finally, we see that these different components of the motion are under independent genetic and adaptive control.

Quantitatively, we found that the exploratory motions of {\em C. elegans} are composed of two principle reorientation elements: abrupt events, including classical $\Omega$--turns, which occur at infrequently,  and the continuous dynamics of orientation between these events.  These two processes both make significant contributions to the worm's total loss of orientational memory on the $\sim 10\,{\rm s}$ time scale.  By focusing on the continuous intervals between discrete turns, however, we find that the worm's orientation can exhibit a longer term memory, lasting two minutes or more, and anti--correlations,  corresponding to an abundance of arcs as seen, for example, in the trajectory of Fig.~1a.  The presence of these arcs as well as the differential behavior of the mutants suggests that {\em C. elegans}   controls more aspects of its motion than the stochastic rate of abrupt events.  Evidence for this sort of continuous control has also recently been observed during {\em C. elegans} chemotaxis \cite{iino_09}.

Extending previous work showing that the body shapes of {\em C. elegans} are captured by four modes \cite{stephens+al_08}, we built a phenomenological model that connects the intrinsic dynamics of these modes to the speed and curvature of the worm's trajectory through the external world.   This model allows us to connect the body configurations, which are what the neuromuscular system can control, to the  behaviors that have adaptive value.  
As an example of what can be learned from this analysis, we studied 
the motion of two mutants, {\em dop-2} and {\em dop-3}, which contain defective receptors for the neurotransmitter dopamine, an important component in the modulation of foraging strategy.  Although these mutants have nearly identical statistics when we look at their discrete turning behaviors, their continuous motions, as seen in the dynamics of fluctuations along four different shape dimensions, show substantial differences.  This suggests that the goal of mapping genes to behavior \cite{brenner_73, brenner_74} will require us to look much more closely, and quantitatively, at
the behavior of individual organisms.

\acknowledgments{We thank T Mora, G Tka\v{c}ik and S N{\o}rrelykke for discussions.
This work was supported in part by NIH grants P50 GM071508 and R01 EY017210, by NSF grant  PHY--0650617,  and by the Swartz Foundation.}

\end{document}